\documentclass[twocolumn,showpacs,preprintnumbers]{revtex4}
\usepackage{graphicx}
\usepackage{dcolumn}
\usepackage{bm}

\begin{document}

\title{Detecting correlation functions of ultracold atoms through Fourier sampling
of time-of-flight images}
\author{L.-M. Duan}
\address{FOCUS center and MCTP, Department of Physics,
University of Michigan, Ann Arbor, MI 48109 }

\begin{abstract}
We propose a detection method for ultracold atoms which allows
reconstruction of the full one-particle and two-particle
correlation functions from the measurements. The method is based
on Fourier sampling of the time-of-flight images through two
consecutive impulsive Raman pulses. For applications of this
method, we discuss a few examples, including detection of phase
separation between superfluid and Mott insulators, various types
of spin or superfluid orders, entanglement, exotic or fluctuating
orders. 
\end{abstract}

\maketitle

Ultracold atomic gas provides an ideal system for controlled study of
various kinds of strongly correlated many-body physics \cite{1,2,3}. To
reveal different many-body phenomena, the correlation function plays a
critical role. In condensed matter systems, information about the
correlation function is typically collected through linear response or
scattering experiments \cite{4}. For ultracold atomic gas, the most powerful
detection technique is arguably the time-of-flight imaging \cite{5}. For
ballistic expansion of the atomic gas, the time-of-flight images actually
give information of the atom number distribution in the momentum space,
which corresponds to the diagonal terms of the one-particle correlation
function. As we have only diagonal correlation function in the momentum
space, it is in general inadequate to use it to reconstruct the correlation
function in the real space.

In this paper, we propose a detection method which extends the
time-of-flight imaging technique by gaining much more information of the
correlation function. The method detects all the non-diagonal correlation
terms as well as the diagonal ones in the momentum space through application
of two consecutive impulsive Raman pulses at the beginning of the expansion
which introduces a tunable momentum difference to the correlation terms.
With this kind of Fourier sampling in the momentum space, one can
reconstruct the full one-particle correlation function in the real space. If
one repeats the time-of-flight imaging measurements many times, as has been
shown in recent works \cite{6,7}, it is also possible to detect the noise
spectroscopy by looking at the statistical correlation between different
images. With a combination of the Fourier sampling and the noise
spectroscopy, one can also reconstruct the full two-particle correlation
function. The reconstruction of the full one-particle and two-particle
correlation functions represents an unprecedented detection ability, which
gives very detailed information of the underlying many-body system. As
illustration of applications, we discuss a few examples, including detection
of phase separation between the superfluid and the Mott insulator states for
bosonic atoms in a trap, various types of spin or superfluid orders in
multi-component Bose or fermi gas, entanglement in an optical lattice,
patterns of the valence bonds associated with some exotic quantum phases
\cite{10}, and the fluctuating orders which exist only in short distance
\cite{11}.

In the time-of-flight imaging experiment, what one measures from light
absorption is the column integrated density of the expanding atomic cloud.
The signal is proportional to the column average of the atom density
operator, which, for the spin component $\alpha $, is denoted as $%
\left\langle n_{\alpha }\left( \mathbf{r,}t\right) \right\rangle
=\left\langle \Psi _{\alpha }^{\dagger }\left( \mathbf{r,}t\right) \Psi
_{\alpha }\left( \mathbf{r,}t\right) \right\rangle $, where $\Psi _{\alpha
}\left( \mathbf{r,}t\right) $ represents the field operator for bosonic or
fermionic atoms at the position $\mathbf{r}$ and the expansion time $t$. We
assume ballistic expansions with the atomic collision effect negligible
during the time of flight \cite{note1}. In such a case, if the expansion
time $t$ is long enough so that the size of the final expanded cloud is much
larger than the size of the initial one, the density $\left\langle n_{\alpha
}\left( \mathbf{r,}t\right) \right\rangle $ is connected with the initial
momentum distribution of the atoms by a simple relation $\left\langle
n_{\alpha }\left( \mathbf{r,}t\right) \right\rangle \propto \left\langle
n_{\alpha }\left( \mathbf{k}\right) \right\rangle =\left\langle \Psi
_{\alpha }^{\dagger }\left( \mathbf{k}\right) \Psi _{\alpha }\left( \mathbf{k%
}\right) \right\rangle $ with the corresponding wave vector $\mathbf{k=}m%
\mathbf{r/}\left( \hbar t\right) $ \cite{5,6}. So the conventional
time-of-flight imaging measures the diagonal terms of the one-particle
correlation function in the momentum space.

To reconstruct the full one-particle correlation function, it is required to
measure also the non-diagonal correlation terms $\left\langle \Psi _{\alpha
}^{\dagger }\left( \mathbf{k}\right) \Psi _{\alpha }\left( \mathbf{k}%
^{\prime }\right) \right\rangle $ in the momentum space. For that purpose,
we propose a detection method as illustrated in Fig. 1, which combines the
time-of-flight imaging with two consecutive impulsive Raman pulses. We
assume there is an additional atomic hyperfine spin level $\beta $ which is
initially empty (such a level is always available for typical alkali atoms).
Right after turnoff of the trapping potential but before any significant
expansion of the atomic cloud, we apply two impulsive Raman pulses to all
the atoms. The duration $\delta t$ of the Raman pulses is short so that one
can neglect the cloud expansion and the atomic collision within $\delta t$.
We assume that the two travelling-wave beams for the first Raman operation
are propagating along different directions with the corresponding wave
vectors $\mathbf{k}_{1}$ and $\mathbf{k}_{2}$, so the effective Raman Rabi
frequency has a spatially varying phase with the form $\Omega \left( \mathbf{%
r}\right) =\Omega _{0}e^{i\left( \delta \mathbf{k\cdot r+\varphi }%
_{1}\right) }$, where $\delta \mathbf{k\equiv k}_{2}-\mathbf{k}_{1}$ and $%
\mathbf{\varphi }_{1}$ is a constant phase. The Hamiltonian within the
interval $\delta t$ can then be written as $H=\int d^{3}\mathbf{r}\Omega
\left( \mathbf{r}\right) \Psi _{\alpha }^{\dagger }\left( \mathbf{r}\right)
\Psi _{\beta }\left( \mathbf{r}\right) +H.c.$ Transferring this Hamiltonian
into the momentum space, we have
\begin{equation}
H=\sum_{\mathbf{k}}\Omega _{0}e^{i\mathbf{\varphi }_{1}}\Psi _{\alpha
}^{\dagger }\left( \mathbf{k}\right) \Psi _{\beta }\left( \mathbf{k+}\delta
\mathbf{k}\right) +H.c.
\end{equation}
We choose the intensity of the Raman beams so that $\Omega _{0}\delta t=\pi
/4$. After this Raman operation, the final field operators in the momentum
space, denoted as $\Psi _{\alpha }^{\prime }\left( \mathbf{k}\right) $ and $%
\Psi _{\beta }^{\prime }\left( \mathbf{k}\right) $, are connected with the
initial ones $\Psi _{\alpha }\left( \mathbf{k}\right) ,\Psi _{\beta }\left(
\mathbf{k}\right) $ through the relation
\begin{eqnarray}
\Psi _{\alpha }^{\prime }\left( \mathbf{k}\right)  &=&\left[ \Psi _{\alpha
}\left( \mathbf{k}\right) +e^{i\mathbf{\varphi }_{1}}\Psi _{\beta }\left(
\mathbf{k+}\delta \mathbf{k}\right) \right] /\sqrt{2},  \nonumber \\
\Psi _{\beta }^{\prime }\left( \mathbf{k}\right)  &=&\left[ \Psi _{\beta
}\left( \mathbf{k}\right) -e^{-i\mathbf{\varphi }_{1}}\Psi _{\alpha }\left(
\mathbf{k-}\delta \mathbf{k}\right) \right] /\sqrt{2}.
\end{eqnarray}
Then, we immediately apply the second Raman operation which is from two
co-propagating laser beams connecting the levels $\alpha ,\beta $ with a
spatially constant effective Rabi frequency $\Omega _{0}e^{i\mathbf{\varphi }%
_{2}}$. With the same pulse area $\Omega _{0}\delta t=\pi /4$, the second
Raman operation induces the transformation
\begin{eqnarray}
\Psi _{\alpha }^{\prime \prime }\left( \mathbf{k}\right)  &=&\left[ \Psi
_{\alpha }^{\prime }\left( \mathbf{k}\right) +e^{i\mathbf{\varphi }_{2}}\Psi
_{\beta }^{\prime }\left( \mathbf{k}\right) \right] /\sqrt{2},  \nonumber \\
\Psi _{\beta }^{\prime \prime }\left( \mathbf{k}\right)  &=&\left[ \Psi
_{\beta }^{\prime }\left( \mathbf{k}\right) -e^{-i\mathbf{\varphi }_{2}}\Psi
_{\alpha }^{\prime }\left( \mathbf{k}\right) \right] /\sqrt{2},
\end{eqnarray}
where $\Psi _{\alpha }^{\prime \prime }\left( \mathbf{k}\right) ,\Psi
_{\beta }^{\prime \prime }\left( \mathbf{k}\right) $ denote the final field
operators after the second Raman operation. We then perform the conventional
time-of-flight imaging, with the atoms in different spin components $\alpha
,\beta $ separated during the flight through a magnetic field gradient \cite
{5}. This imaging measures the momentum distribution $\left\langle n_{\alpha
}^{\prime \prime }\left( \mathbf{k}\right) \right\rangle =\left\langle \Psi
_{\alpha }^{\prime \prime \dagger }\left( \mathbf{k}\right) \Psi _{\alpha
}^{\prime \prime }\left( \mathbf{k}\right) \right\rangle $ and $\left\langle
n_{\beta }^{\prime \prime }\left( \mathbf{k}\right) \right\rangle
=\left\langle \Psi _{\alpha }^{\prime \prime \dagger }\left( \mathbf{k}%
\right) \Psi _{\alpha }^{\prime \prime }\left( \mathbf{k}\right)
\right\rangle $. With a combination of Eqs. (2) and (3), we find that the
difference between the two images $\left\langle n_{\alpha \beta }^{\prime
\prime }\left( \mathbf{k}\right) \right\rangle \equiv \left\langle n_{\alpha
}^{\prime \prime }\left( \mathbf{k}\right) \right\rangle -\left\langle
n_{\beta }^{\prime \prime }\left( \mathbf{k}\right) \right\rangle $ gives
the non-diagonal correlation terms
\begin{equation}
\left\langle n_{\alpha \beta }^{\prime \prime }\left( \mathbf{k}\right)
\right\rangle =-\mathop{\rm Re}\nolimits\left\langle \Psi _{\alpha
}^{\dagger }\left( \mathbf{k}\right) \Psi _{\alpha }\left( \mathbf{k-}\delta
\mathbf{k}\right) e^{i\delta \mathbf{\varphi }}\right\rangle ,
\end{equation}
where $\delta \mathbf{\varphi =\varphi }_{2}-\mathbf{\varphi }_{1}$ and we
have used the fact that $\Psi _{\beta }\left( \mathbf{k}\right) $ is
initially in the vacuum state.

\begin{figure}[tbph]
\centering
\includegraphics[height=3cm,width=8cm]{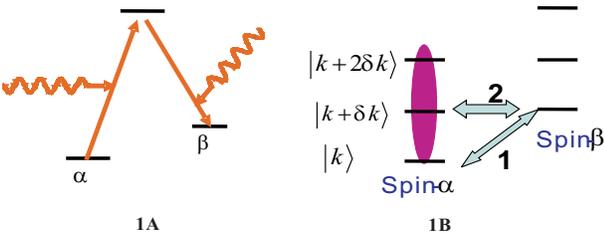}
\caption{ (1A) Raman pulses with different propagating directions
introduce a momentum kick. (1B) The momentum ladder connected by
two consecutive impulsive Raman pulses. The first $\protect\pi/2$
Raman pulse (pulse $1$) introduces a tunable momentum kick
$\protect\delta k$, while the second one (pulse $2$) has no kick.
The atoms are initially in the spin-$\alpha$ state.} \label{Fig1}
\end{figure}

With the above method, we have shown how to measure the real and imaginary
parts of the non-diagonal correlation function $\left\langle \Psi _{\alpha
}^{\dagger }\left( \mathbf{k}\right) \Psi _{\alpha }\left( \mathbf{k-}\delta
\mathbf{k}\right) \right\rangle $ by choosing the relative phase $\delta
\mathbf{\varphi =0}$ or $\pi /2$, respectively. The momentum difference $%
\delta \mathbf{k}$ is controlled by the relative angle of the two laser
pulses for the first Raman operation. By varying this angle, $\delta \mathbf{%
k}$ can be varied from $0$ to $2\mathbf{k}_{0}$, where $\left| \mathbf{k}%
_{0}\right| =2\pi /\lambda $ is the optical wave vector. With a Fourier
transform of $\left\langle \Psi _{\alpha }^{\dagger }\left( \mathbf{k}%
\right) \Psi _{\alpha }\left( \mathbf{k-}\delta \mathbf{k}\right)
\right\rangle $, one can reconstruct the real-space correlation function $%
\left\langle \Psi _{\alpha }^{\dagger }\left( \mathbf{r+}\delta \mathbf{r/2}%
\right) \Psi _{\alpha }\left( \mathbf{r-}\delta \mathbf{r/2}\right)
\right\rangle =\int \left\langle \Psi _{\alpha }^{\dagger }\left( \mathbf{k+}%
\delta \mathbf{k/2}\right) \Psi _{\alpha }\left( \mathbf{k-}\delta \mathbf{%
k/2}\right) \right\rangle e^{i\left( \mathbf{k\cdot }\delta \mathbf{r+\delta
k\cdot r}\right) }d\delta \mathbf{k}d\mathbf{k}/2\pi $, with a spatial
resolution in $\mathbf{r}$ down to $2\pi /\left| \delta \mathbf{k}\right|
_{\max }=\lambda /2$. This kind of Fourier sampling in the momentum space
through the laser phase gradient allows us to directly probe very small
spatial structure in the ultracold atomic gas, although the probe beams are
always shined on all the atoms without separate addressing of any particular
region. For instance, for the case of atoms in an optical lattice, with such
a resolution, we can reconstruct the correlation $\left\langle a_{\alpha
}^{\dagger }\left( i\right) a_{\alpha }\left( j\right) \right\rangle $ of
the mode operators $a_{\alpha }$ at two arbitrary sites $i$ and $j$ (the
lattice spacing is $\lambda /2$).

Before showing how to measure the full two-particle correlation function, we
add a few remarks here. First, note that in the above method, it is
essential to apply two consecutive Raman operations. If we introduce
momentum difference $\delta \mathbf{k}$ by applying only one Raman operation
on the same spin component $\alpha $ with the Hamiltonian $H^{\prime }=\sum_{%
\mathbf{k}}\Omega _{0}e^{i\mathbf{\varphi }_{1}}\Psi _{\alpha }^{\dagger
}\left( \mathbf{k}\right) \Psi _{\alpha }\left( \mathbf{k+}\delta \mathbf{k}%
\right) +H.c.$, the $H^{\prime }$ will couple the whole momentum ladder $%
\Psi _{\alpha }\left( \mathbf{k}\right) $, $\Psi _{\alpha }\left( \mathbf{%
k\pm }\delta \mathbf{k}\right) $, $\Psi _{\alpha }\left( \mathbf{k\pm 2}%
\delta \mathbf{k}\right) $, $\cdots $, and we thus cannot get a simple form
of transformation such as those given by Eqs. (2) and (3). So the method
with two consecutive Raman operations seems to be the easiest one for
measuring the non-diagonal momentum correlations. Second, in the above
method, we have assumed to tune the momentum difference by changing the
relative angle of two laser pulses. This tuning can also be achieved by
applying a sequence of laser pulses incident along two fixed directions with
a small angle $\delta \theta $, with each pair of pulses introducing a fixed
momentum kick $\mathbf{k}_{\delta }=2\left| \mathbf{k}_{0}\right| \sin
\left( \delta \theta /2\right) \simeq \left| \mathbf{k}_{0}\right| \delta
\theta $ (see illustration in Fig. 2 and its caption). With a maximum of $N$
such pulse pairs, the momentum difference $\delta \mathbf{k}$ in the
correlation function (4) can be tuned among different values $\mathbf{k}%
_{\delta }$, $2\mathbf{k}_{\delta },$ $\cdots $, $N\mathbf{k}_{\delta }$,
which corresponds to a discrete Fourier sampling of the atomic correlation
function with the momentum-space and real-space resolutions given
respectively by $\left| \mathbf{k}_{0}\right| \delta \theta $ and $\lambda
/\left( N\delta \theta \right) $. Finally, the above method can be
generalized straightforwardly to measure also the spin-spatial correlations.
If we have multiple Zeeman spin components $\alpha ,\alpha ^{\prime },\cdots
$ for the state of the atomic cloud, one can reconstruct the full
spin-spatial correlation $\left\langle \Psi _{\alpha }^{\dagger }\left(
\mathbf{k}_{\mathbf{1}}\right) \Psi _{\alpha ^{\prime }}\left( \mathbf{k}%
_{2}\right) \right\rangle $ (or $\left\langle \Psi _{\alpha }^{\dagger
}\left( \mathbf{r}_{\mathbf{1}}\right) \Psi _{\alpha ^{\prime }}\left(
\mathbf{r}_{2}\right) \right\rangle $) through a combination of the above
Fourier sampling with a pair of Raman pulses which mixes the spin components
$\alpha ,\alpha ^{\prime }$.

\begin{figure}[tbph]
\centering
\includegraphics[height=3cm,width=8cm]{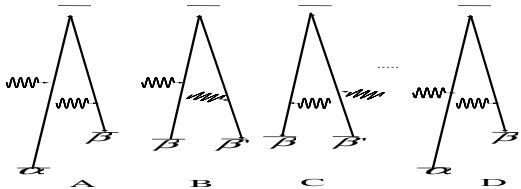}
\caption{ Tunable momentum kicks with pulses propagating in fixed
directions. Steps (A) and (D) represent $\protect\pi/2$ pulses
with no momentum kicks. Steps (B) and (C) represent $\protect\pi$
pulses which a small fixed momentum kick $k_{\protect\delta}$
(pulses for B, C are propagating in reverse directions). The
intermediate pulses B, C can be applied with a controlled number
of times, which correspond a discrete Fourier sampling of the
correlation function with a momentum-space resolution of
$k_{\protect\delta}$.} \label{Fig2}
\end{figure}

The above method, combined with the noise spectroscopy, can also be used to
reconstruct the full two-particle correlation function $\left\langle \Psi
_{\alpha }^{\dagger }\left( \mathbf{r}_{1}\right) \Psi _{\alpha }^{\dagger
}\left( \mathbf{r}_{2}\right) \Psi _{\alpha }\left( \mathbf{r}_{3}\right)
\Psi _{\alpha }\left( \mathbf{r}_{4}\right) \right\rangle $. For the noise
spectroscopy \cite{6,7}, one just needs to note that for different runs of
experiment, the time-of-flight images can have quantum fluctuation even if
one starts with the same state of the atomic cloud. One image corresponds to
a single-run measurement of the column integrated atomic density operator,
and by looking at correlation of different images, one can find out quantum
correlation of the density operator. To detect such quantum correlation, all
the technical noise for the images needs to be reduced below the level of
quantum noise. It is remarkable that the recent experiments have shown that
such quantum correlation of images is indeed visible after suppression of
the technical noise \cite{7}. Each time-of-flight imaging corresponds to a
measurement of the one-particle momentum correlation $\left\langle \Psi
_{\alpha }^{\dagger }\left( \mathbf{k}\right) \Psi _{\alpha }\left( \mathbf{k%
}-\delta \mathbf{k}\right) \right\rangle .$ To reconstruct the full
two-particle correlation function, one just needs to repeat each of such
imaging measurements $M$ times ($M$ needs to be sufficiently large so that $%
1/\sqrt{M}$, which characterize the statistical error for the quantum
correlation measurement, is sufficiently small). Then, by looking at the
correlation of arbitrary two images corresponding to $\left\langle \Psi
_{\alpha }^{\dagger }\left( \mathbf{k}\right) \Psi _{\alpha }\left( \mathbf{k%
}-\delta \mathbf{k}\right) \right\rangle $ and $\left\langle \Psi _{\alpha
}^{\dagger }\left( \mathbf{k}^{\prime }\right) \Psi _{\alpha }\left( \mathbf{%
k}^{\prime }-\delta \mathbf{k}^{\prime }\right) \right\rangle $
respectively, one gets the two-particle correlation $\left\langle \Psi
_{\alpha }^{\dagger }\left( \mathbf{k}\right) \Psi _{\alpha }\left( \mathbf{k%
}-\delta \mathbf{k}\right) \Psi _{\alpha }^{\dagger }\left( \mathbf{k}%
^{\prime }\right) \Psi _{\alpha }\left( \mathbf{k}^{\prime }-\delta \mathbf{k%
}^{\prime }\right) \right\rangle $. As the two images can be arbitrarily
chosen (with different $\mathbf{k,k}^{\prime },\delta \mathbf{k,}\delta
\mathbf{k}^{\prime }$), this gives us the full two-particle correlation in
the momentum space, and the real-pace correlation $\left\langle \Psi
_{\alpha }^{\dagger }\left( \mathbf{r}_{1}\right) \Psi _{\alpha }\left(
\mathbf{r}_{2}\right) \Psi _{\alpha }^{\dagger }\left( \mathbf{r}_{3}\right)
\Psi _{\alpha }\left( \mathbf{r}_{4}\right) \right\rangle $ is simply its
Fourier transform. The spatial resolution in this case is still given by $%
\lambda /2$ similar to the one-particle correlation measurement.

The measurement of the full one-particle and two-particle correlation
functions gives very detailed information of the many-body properties of the
underlying system. Here, we give a few examples for illustration of its
applications. We start with one-particle correlation function. As the first
example, note that for the superfluid to Mott insulator transition in a
global harmonic trap (as it is the case for experiments), theory has
predicted phase separation with layers of superfluid states intervened in by
layers of Mott insulator states with decreasing integer filling numbers \cite
{9}. Such a picture have not been confirmed yet by experiments as with the
conventional time-of-flight imaging, it is hard to see this layer-by-layer
structure. Through the Fourier sampling, however, one can reconstruct the
one-particle correlation $\left\langle \Psi _{\alpha }^{\dagger }\left(
\mathbf{r}_{1}\right) \Psi _{\alpha }\left( \mathbf{r}_{2}\right)
\right\rangle $. From this correlation, it would be evident to see the phase
separation: the Mott insulator state is characterized by a vanishing
non-diagonal correlation and a constant (integer) diagonal correlation,
while the superfluid state is characterized by both non-zero diagonal and
non-diagonal correlations which vary continuously in space.

With ultracold atoms in an optical lattice, one can realize different
magnetic Hamiltonians \cite{13}. Depending on the lattice geometry and
interaction configurations, such Hamiltonians may support various types of
magnetic orders \cite{13,14}. The magnetic orders typically can be written
in the form $\left\langle \mathbf{S}_{i}\right\rangle =\mathbf{v}_{1}\cos
\left( \mathbf{K\cdot r}_{i}\right) +\mathbf{v}_{2}\sin \left( \mathbf{%
K\cdot r}_{i}\right) $ \cite{10}, where $\mathbf{S}_{i}$ is the spin
operator on the site $i$ with the coordinate $\mathbf{r}_{i}$, $\mathbf{v}%
_{1}$ and $\mathbf{v}_{2}$ are two vectors specifying certain directions,
and $\mathbf{K}$ characterizes spatial variation of the order parameter. The
spin $\mathbf{S}_{i}$ is defined as $\mathbf{S}_{i}=\sum_{\alpha \alpha
^{\prime }}a_{i\alpha }^{\dagger }\mathbf{\sigma }_{\alpha \alpha ^{\prime
}}a_{i\alpha ^{\prime }}/2$, where $\mathbf{\sigma }$ is the Pauli matrix
and $a_{i\alpha }$ is the mode operator which is connected with the field
operator $\Psi _{\alpha }\left( \mathbf{r}\right) $ through $a_{i\alpha
}=\int w_{i}^{\ast }\left( \mathbf{r}\right) \Psi _{\alpha }\left( \mathbf{r}%
\right) d^{3}\mathbf{r}$ ($w_{i}^{\ast }\left( \mathbf{r}\right) $ is the
Wannier function for the site $i$). Through Fourier sampling of the
time-of-flight images with a resolution $\lambda /2$, one can reconstruct
correlation of the mode operators $\left\langle a_{i\alpha }^{\dagger
}a_{j\alpha ^{\prime }}\right\rangle $ on any sites ($i=j$ for the special
case). So one can directly detect any type of spatial variation of the order
parameter $\left\langle \mathbf{S}_{i}\right\rangle $ (note that the
conventional time-of-flight imaging cannot detect spatially varying order
parameters).

Another interesting application of this Fourier sampling technique is to
detect local fluctuating orders. Local fluctuating orders, such as stripes
(unidirectional density waves), have received wide attention in strongly
correlated physics (in particular for high-Tc superconductors) \cite{11}.
The local fluctuating order typically takes place near a critical point with
competing orders or at the proximity of an ordered phase. With ultracold
fermions in an optical lattice, the fundamental Hamiltonian is very similar
to those models for high-Tc superconductors \cite{15,16}. One expects that
competing fluctuating orders may arise as well in the phase diagram of this
system. The local fluctuating orders such as stripes correspond to
real-space patterns of some micro-phase separation \cite{11}, and the
conventional time-of-flight imaging cannot see it because it will be
averaged out. However, through the Fourier sampling technique with a spatial
resolution down to the lattice constant, it should be evident to detect
these local fluctuating orders whenever they show up. A measurement of the
one-particle correlation $\left\langle \Psi _{\alpha }^{\dagger }\left(
\mathbf{r}_{1}\right) \Psi _{\alpha ^{\prime }}\left( \mathbf{r}_{2}\right)
\right\rangle $ is typically enough to probe the fluctuating orders such as
the local density waves (stripes).

The above discussion shows some applications of the one-particle
correlation. With the two-particle correlation measured, one can gain
further information. For example, one can use the two-particle correlation
to unambiguously detect the entanglement pattern for ultracold atoms in an
optical lattice. Through controlled atomic collisions, one can generate
entanglement for atoms on different lattice sites \cite{17}, and some
initial experimental evidence of entanglement has been reported for such a
system \cite{18}. It is hard however to directly detect atomic entanglement
between different sites or regions inside an optical lattice as the
conventional detection technique donot have such a spatial resolution. This
entanglement can be unambiguously confirmed with a measurement of spin
correlations for atoms on different sites (similar to the Bell inequality
measurement). As each spin operator involves only two atomic mode operators,
the spin correlation is included in the two-particle correlation function.
So, a measurement of the two-particle correlation can gives information of
spin entanglement pattern in the lattice.

For fermionic systems with pairing instability (such as the fermionic
superfluid state \cite{3}), it is also desirable to have a method to
directly measure the Cooper pair function $\left\langle \Psi _{\alpha
}\left( \mathbf{r}_{1}\right) \Psi _{\alpha ^{\prime }}\left( \mathbf{r}%
_{2}\right) \right\rangle .$ The two-particle correlation also gives
information of such pair function as with Cooper pairing, the two-particle
correlation can be typically approximated with the following decomposition $%
\left\langle \Psi _{\alpha }^{\dagger }\left( \mathbf{r}_{1}\right) \Psi
_{\alpha ^{\prime }}^{\dagger }\left( \mathbf{r}_{2}\right) \Psi _{\alpha
^{\prime }}\left( \mathbf{r}_{3}\right) \Psi _{\alpha }\left( \mathbf{r}%
_{4}\right) \right\rangle \approx \left\langle \Psi _{\alpha }^{\dagger
}\left( \mathbf{r}_{1}\right) \Psi _{\alpha ^{\prime }}^{\dagger }\left(
\mathbf{r}_{2}\right) \right\rangle \left\langle \Psi _{\alpha ^{\prime
}}\left( \mathbf{r}_{3}\right) \Psi _{\alpha }\left( \mathbf{r}_{4}\right)
\right\rangle +\left\langle \Psi _{\alpha }^{\dagger }\left( \mathbf{r}%
_{1}\right) \Psi _{\alpha }\left( \mathbf{r}_{4}\right) \right\rangle
\newline
\left\langle \Psi _{\alpha ^{\prime }}^{\dagger }\left( \mathbf{r}%
_{2}\right) \Psi _{\alpha ^{\prime }}\left( \mathbf{r}_{3}\right)
\right\rangle -\left\langle \Psi _{\alpha }^{\dagger }\left( \mathbf{r}%
_{1}\right) \Psi _{\alpha ^{\prime }}\left( \mathbf{r}_{3}\right)
\right\rangle \left\langle \Psi _{\alpha ^{\prime }}^{\dagger
}\left( \mathbf{r}_{2}\right) \Psi _{\alpha }\left(
\mathbf{r}_{4}\right) \right\rangle $ by applying the Wick
theorem. As the one-particle correlation is known already with the
Fourier sampling, so a measurement of the two-particle correlation
gives the pair function $\left\langle \Psi _{\alpha }\left(
\mathbf{r}_{1}\right) \Psi _{\alpha ^{\prime }}\left(
\mathbf{r}_{2}\right) \right\rangle $ (the two-particle
correlation actually contains more information, for instance, it
can be used to check self-consistently whether the above
decomposition is valid). The pair function $\left\langle \Psi
_{\alpha }\left( \mathbf{r}_{1}\right) \Psi _{\alpha ^{\prime
}}\left( \mathbf{r}_{2}\right) \right\rangle $ can tell us the
symmetry (s-wave or d-wave for instance) and the size of the
Cooper pairs as well as how the pair structure changes in space.

Finally, a potentially more interesting application of the two-particle
correlation might be that it gives us a way to directly detect the patterns
of valence bonds in an optical lattice. A valence bond on the sites $i,j$
(not necessarily neighbors) is simply the ground state of the bond operator $%
Q_{ij}=\mathbf{S}_{i}\cdot \mathbf{S}_{j}$ ($\mathbf{S}_{i}$ are spin
operators) \cite{10}. The resonating valence bond (RVB) states \cite{19,10},
which include many possible patterns of the valence bond distribution in the
lattice, have been conjectured as one of the most likely ground state of the
$t$-$J$ or Hubbard models in the strongly correlated limit. As the number of
possible configurations of the RVB states increses exponentially with the
size of the lattice, it is hard to figure out the distribution pattern of
the valence bonds in an lattice. However, with an atomic realization of the $%
t$-$J$ model \cite{16}, one can directly measure the bond operators $Q_{ij}$
in an optical lattice to find out the most likely distribution pattern of
the valence bonds. Each bond operator $Q_{ij}$ corresponds to a special
component of the two-particle correlation function, so a measurement of the
two-particle correlation with Fourier sampling gives complete information
about the bond distribution.

In summary, we have proposed a detection method which combines the
commonly-used time-of-flight imaging technique with the Fourier sampling
based on application of two consecutive impulsive Raman pulses. This
detection method allows us to reconstruct the full one-particle and
two-particle correlation functions. We have discussed a few examples for
illustration of the wide applications of the correlation functions for
probing many-body properties of the underlying system.

This work was supported by the NSF award (0431476), the ARDA under ARO
contracts, and the A. P. Sloan Fellowship.


\begin{thebibliography}{99}
\bibitem{1}  For a review, see D. Jaksch, P. Zoller, cond-mat/0410614.

\bibitem{2}  M. Greiner, et al., Nature 415, 39 (2002); C. Orzel, et al.,
Science 291, 2386 (2001).

\bibitem{3}  C.A. Regal, M. Greiner and D.S. Jin, Phys. Rev. Lett. \textbf{92%
}, 040403 (2004); M.W. Zwierlein \textit{et al.}, Phys. Rev. Lett. \textbf{92%
}, 120403 (2004); C. Chin \textit{et al.}, Science \textbf{305}, 1128
(2004); J. Kinast \textit{et al.}, Science \textbf{307}, 1296 (2005); M.W.
Zwierlein \textit{et al.}, Nature 435, 1047 (2005); M. Holland \textit{et al.%
}, Phys. Rev. Lett. \textbf{87}, 120406 (2001).

\bibitem{4}  W. Jones, and N. H. March, \textit{Theoretical solid state
physics}, Dove, New York (1985).

\bibitem{5}  For a review, see W. Ketterle, D.S. Durfee, D.M. Stamper-Kurn,
cond-mat/9904034.

\bibitem{6}  E. Altman, E. Demler, M. D. Lukin, Phys. Rev. A 70, 013603
(2004).

\bibitem{7}  S. Folling et al., Nature 434, 481 (2005); M. Greiner et al.,
Phys. Rev. Lett. 94, 110401 (2005); C.-S. Chuu et al., quant-ph/0508143.

\bibitem{9}  D. Jaksch, et al., Phys. Rev. Lett. \textbf{81}, 3108 (1998);
B. DeMarco et al., cond-mat/0501718.

\bibitem{10}  S. Sachdev, Rev. Mod. Phys. 75, 913 (2003).

\bibitem{11}  S.A.Kivelson et al., Rev. Mod. Phys. 75, 1201 (2003).

\bibitem{note1}  For optical lattice experiments with small filling factors,
the ballistic expansion is typically a good approximation. With the Feshbach
resonance technique, one can also always tune the collision interaction to a
negligible magnitude at the beginning of the expansion to satisfy the
condition of the ballistic expansion.


\bibitem{13}  L.-M. Duan, E. Demler, M. D. Lukin, Phys. Rev. Lett. 91,
090402 (2003).

\bibitem{14}  L. Santos et al., Phys. Rev. Lett. 93, 030601 (2004).

\bibitem{15}  W. Hofstetter et al., Phys. Rev. Lett. 89, 220407 (2002).

\bibitem{16}  L.-M. Duan, cond-mat/0508745.

\bibitem{17}  D. Jaksch, et al., Phys.Rev.Lett. 82, 1975 (1999).

\bibitem{18}  O. Mandel et al., Nature 425, 937 (2003).

\bibitem{19}  P. W. Anderson, Science 235, 1196 (1987).
\end{thebibliography}
\end{document}